\documentclass[aps,prl,floatfix,superscriptaddress,reprint]{revtex4-2}

\usepackage{graphicx}
\usepackage{color}
\usepackage{amsfonts}
\usepackage{amssymb}
\usepackage{amsmath}
\usepackage{hyperref}
\usepackage{cleveref}
\usepackage{braket}
\usepackage{mathrsfs}

\newcommand{\mc}[1]{\mathcal{ #1}} 

\definecolor{green}{rgb}{0, 1, 0}

\begin{document}

\title{Observable-enriched entanglement}

\author{Joe H. Winter}
\affiliation{Max Planck Institute for Chemical Physics of Solids, N{\"o}thnitzer Stra{\ss}e 40, 01187 Dresden, Germany}
\affiliation{Max Planck Institute for the Physics of Complex Systems, N{\"o}thnitzer Stra{\ss}e 38, 01187 Dresden, Germany}
\affiliation{SUPA, School of Physics and Astronomy, University of St.\ Andrews, North Haugh, St.\ Andrews KY16 9SS, UK}

\author{Reyhan Ay}
\affiliation{Max Planck Institute for Chemical Physics of Solids, N{\"o}thnitzer Stra{\ss}e 40, 01187 Dresden, Germany}
\affiliation{Max Planck Institute for the Physics of Complex Systems, N{\"o}thnitzer Stra{\ss}e 38, 01187 Dresden, Germany}
\affiliation{Izmir Institute of Technology, Gülbahçe Kampüsü, 35430 Urla Izmir, Türkiye}

\author{Bernd Braunecker}
\affiliation{SUPA, School of Physics and Astronomy, University of St.\ Andrews, North Haugh, St.\ Andrews KY16 9SS, UK}

\author{A. M. Cook}
\affiliation{Max Planck Institute for Chemical Physics of Solids, N{\"o}thnitzer Stra{\ss}e 40, 01187 Dresden, Germany}
\affiliation{Max Planck Institute for the Physics of Complex Systems, N{\"o}thnitzer Stra{\ss}e 38, 01187 Dresden, Germany}

\begin{abstract}
We introduce methods of characterizing entanglement, in which entanglement measures are enriched by the
matrix representations of operators for observables. These observable
operator matrix representations can enrich the partial trace over subsets of a
system's degrees of freedom, yielding reduced density matrices useful in
computing various measures of entanglement, which also preserve the observable expectation value. We focus here on applying these
methods to compute \textit{observable-enriched} entanglement spectra, unveiling new bulk-boundary correspondences of canonical four-band models for topological skyrmion phases and their connection to simpler forms of bulk-boundary correspondence. Given the fundamental roles entanglement signatures and observables play in study of
quantum many body  systems, observable-enriched entanglement is broadly applicable to
myriad problems of quantum mechanics.
\end{abstract}

\maketitle

Entanglement is essential in characterising quantum many body systems given its role in quantum information theory~\cite{vedral2002, ohya2004quantum, eisert2006entanglement}, with various measures of entanglement applied for characterising topological states~\cite{kitaev2006, levin2006, fidkowski2010}, quantum critical phenomena~\cite{vidal2003, jin_quantum_2004, PasqualeCalabrese_2004}, phase transitions~\cite{TatsumaNishioka_2007, KLEBANOV2008274, AriPakman_2008}, and dynamics~\cite{Eisler_2007, Calabrese_2007, Calabrese_2016}.  Integral to entanglement characterisation is the partial trace operation~\cite{nielsen2010quantum, preskillnotes, wilde2017quantum}: considerable information about quantum systems derives from  partial trace applied to the density matrix over myriad subsets of Hilbert space~\cite{alexandradinata2011, rodriguez2012, dubail2012, sterdyniak2012}.


Such entanglement measures provide information about the connection between the Hilbert spaces of physical subsystems that goes beyond the decomposition into joint basis states. Given the on-going research into the meaning of these measures, we show that similar information can be encoded in observable representations. To this end, we introduce a generalized, observable enriched (OE) partial trace (OEPT) $\Tilde{\mathrm{Tr}}$ with respect to an observable $S$. When applied on the full density matrix $\rho$, the OEPT produces an auxiliary density matrix $\rho_s = \Tilde{\mathrm{Tr}}[\rho]$, which is constructed entirely from $\langle S \rangle$ and captures all entanglement and topological information by the requirement that $\text{Tr}[\rho S ] = \text{Tr}[\rho_s \tilde{S}]$, where $\text{Tr}$ is the trace operation. That is, the expectation value of the observable is preserved by $\tilde{\text{T}}\text{r}$.





We demonstrate the power of our methods by studying topological Skyrmion phases of matter~\cite{cook2023}, which are lattice counterparts of the quantum skyrmion Hall effect~\cite{qskhe}. Our methods reveal essential features of these topological states, which generalise those~\cite{PhysRevLett.50.1395, FQH-Laughlin, kallin1984excitations, PhysRevLett.62.82, halperin1993theory, halperin1982quantized, halperin1984statistics, wen1991gapless, PhysRevLett.63.199, BERNEVIG2002185, nayak2008, sarma_majorana_2015, freedman_two-eigenvalue_2002, Kitaev_2001} within the framework of the quantum Hall effect (QHE)~\cite{klitzing1980, PhysRevLett.48.1559}. Notably, we find first evidence of the generalised bulk-boundary correspondence of topological Skyrmion states~\cite{qskhe} in the particularly simple four-band Hamiltonians~\cite{liu2023} capturing this physics, and we utilise OE entanglement to reveal it as a generalisation of bulk-boundary correspondence in the QHE framework.



\textit{Topological Skyrmion phases of matter}---We first briefly introduce key concepts of topological Skyrmion phases of matter and the quantum Skyrmion Hall effect to later introduce our methods. The first known topological
Skyrmion phases are $2+1$-dimensional topological states possible in
effectively non-interacting systems~\cite{cook2023, liu2023}, which are
characterized by the topological invariant $\mc{Q}$, the topological charge of
winding in the spin texture of occupied states over the Brillouin zone defined
by momentum components $k_x$ and $k_y$, or
\begin{equation}
    \mc{Q} = {1 \over 4\pi} \int_{BZ}d\boldsymbol{k} \left[ \langle \boldsymbol{\hat{S}}(\boldsymbol{k}) \rangle \cdot \left(\partial_{k_x}\langle \boldsymbol{\hat{S}}(\boldsymbol{k}) \rangle \times \partial_{k_y}\langle \boldsymbol{\hat{S}}(\boldsymbol{k}) \rangle \right) \right],
    \label{skyrmnum}
\end{equation}

where $\boldsymbol{S} = \left(S_1, S_2, S_3\right)$ is the spin representation
and $\langle \boldsymbol{\hat{S}}(\boldsymbol{k}) \rangle = \langle \boldsymbol{S}(\boldsymbol{k}) \rangle/|\boldsymbol{S}|$ is the normalized
expectation value of the spin for occupied states. Here $\langle
S_i(\boldsymbol{k}) \rangle= \sum_{n \in \mathrm{occ}} \langle n,
\boldsymbol{k} | S_i | n, \boldsymbol{k} \rangle $, with $i \in \{1,2,3\}$ and
$| n, \boldsymbol{k} \rangle$ the Bloch state associated with the
$n$\textsuperscript{th} band. For systems with only a spin degree
of freedom (DoF), $\mc{Q}$ is the total Chern number, but decouples from the total Chern number in systems with
multiple DoFs, to characterize a topological state distinct from the Chern
insulator~\cite{cook2023, liu2023, calderon2023_TRIskyrm}.

Defining the full Hamiltonian of the system as $\mc{H}$, where $\mc{H} = \sum_{\boldsymbol{k}} \Psi^{\dagger}_{\boldsymbol{k},\alpha,\beta} H(\boldsymbol{k}) \Psi^{}_{\boldsymbol{k},\alpha,\beta}$  in terms of Bloch Hamiltonian $H(\boldsymbol{k})$, we choose the basis $\Psi^{}_{\boldsymbol{k}} = (c^{}_{\boldsymbol{k},+} c^{}_{\boldsymbol{k},-}, c^{\dagger}_{\boldsymbol{k},-}, c^{\dagger}_{\boldsymbol{k},+})^{T}$, where $c^{}_{\boldsymbol{k},\alpha}$ annihilates a fermion with momentum $\boldsymbol{k}$ and $\alpha \in \{+, - \}$ defines a two-fold (pseudo)spin DoF. For the purposes of this discussion, it is sufficient to consider $\alpha$ as a spin-$1/2$ DoF.

The Bloch Hamiltonian $H(\boldsymbol{k})$ is then a
generalized Bogoliubov de Gennes (BdG) Hamiltonian,
consisting of a generic two band normal state Hamiltonian $H_N(\boldsymbol{k})$
and pairing term $\Delta(\boldsymbol{k})$ as
\begin{equation}
	H_{BdG}(\boldsymbol{k}) = \begin{bmatrix} H_N(\boldsymbol{k}) &
	\Delta(\boldsymbol{k}) \\ \Delta^{\dagger}(\boldsymbol{k}) &
-H^{T}_N(\boldsymbol{k}) \end{bmatrix}
	\label{BdGHam}
.\end{equation}


We consider $H_N(\boldsymbol{k}) = h_{0}(\boldsymbol{k}) \mathbf{I} +
\boldsymbol{h}(\boldsymbol{k}) \cdot \boldsymbol{\sigma} $ and $\Delta(\boldsymbol{k}) = i \Delta_{0} (d_{0}(\boldsymbol{k}) +
\boldsymbol{d}(\boldsymbol{k}) \cdot \boldsymbol{\sigma}) \sigma_{y} $. In
these expressions, $\Delta_{0}$ represents a constant;
$h_{0}(\boldsymbol{k})$ and $d_{0}(\boldsymbol{k})$ are real scalar functions;
$\boldsymbol{h}(\boldsymbol{k})$ and $\boldsymbol{d}(\boldsymbol{k})$ are real
vector functions and finally, $\boldsymbol{\sigma} =
\left( \sigma_x, \sigma_y, \sigma_z \right)$, where $\sigma_{\mu}$ is the
$\mu$\textsuperscript{th} Pauli matrix.\par


\textit{Observable enriched partial trace}--- We now introduce the observable enriched partial trace (OEPT) by mapping the ground state (GS) of $H(k)$, represented by the density matrix $\rho_{GS}$, onto an auxiliary system that reproduces the same expectation values of $S$. To this end we define a two-level density matrix $\rho_s$, for some state of an auxiliary system with Bloch Hamiltonian $h(\boldsymbol{k})$ and basis $\psi_{\boldsymbol{k}} =
( c^{}_{\boldsymbol{k}, +},c^{}_{\boldsymbol{k}, -})^{T}$.
That is, $\psi$ possesses only the spin-$1/2$ DoF for each momentum
$\boldsymbol{k}$. Subsequently, we may enforce the following
relation
\begin{equation}
	\text{Tr}[\rho_{\mathrm{GS}}(\boldsymbol{k}) S_{\mu} ] = \text{Tr}[\rho_{s}(\boldsymbol{k})  \sigma_{\mu}],
	\label{expparttrc}
	\end{equation}
which implicitly defines a $\rho_s(\boldsymbol{k}) $ yielding the same spin
expectation value as $\rho_{\mathrm{GS}}(\boldsymbol{k}) $, despite their
different corresponding spin representations.

Using $SU(2)$ commutation relations and trace properties, \cref{expparttrc} yields $\rho_{s}\left(\boldsymbol{k} \right)$ of the form:
\begin{equation}
    \rho_{s}(\boldsymbol{k}) =  \left( \mathbf{I}_2 + \braket{\boldsymbol{S}(\boldsymbol{k})} \cdot \boldsymbol{\sigma} \right)/2
,\end{equation}
where $\mathbf{I}_2$ is the $2 \times 2$ identity matrix.

 Now consider the existence of a unitary transformation which maps spin representation $\{ S^{\mu} \}_{\mu \in 1,2,3}$  onto $ \{ \mathbf{I}_2 \otimes \sigma_{\mu} \}_{\mu \in
x,y,z}$. Then the spin, $S$, is completely decoupled from the non-spin DoFs, $\overline{S}$, and consequently the map from $\rho_{GS} \to \rho_{s}$ reduces to performing a
partial trace over the non-spin degrees of freedom, $\bar{S}$, as:

\begin{equation}
	\hat{O}_S = \mathbf{I}_{\overline{S}} \otimes \hat{o}_S \implies \text{Tr} \left[ \text{Tr}_{\overline{S}}[\rho] \ \hat{o}_S\right] = \text{Tr} [ \rho \hat{O}_S ]
.\end{equation}

We interpret our system as consisting of a spin DoF coupled to a bath of
the non-spin DoFs which are traced out during the OEPT. The
$\rho_s\left( \boldsymbol{k}\right)$ is then interpreted as a reduced density
operator useful in computing various entanglement properties~\cite{nielsen2010quantum, preskillnotes, wilde2017quantum}. Notably, even if the
spin and particle-hole DoFs are not separable, $\rho_s(\boldsymbol{k})$ remains a valid
density matrix by construction and faithfully reproduces the GS spin
expectation values over the Brillouin zone (BZ). This exemplifies the utility of this technique.

To proceed, we note that the class of Hamiltonians is $\mathscr{C}'$-symmetric, which is a generalised charge conjugation symmetry defined by $ \mathscr{C}'^{-1} H(\boldsymbol{k}) \mathscr{C}' = - H(\boldsymbol{k})^{T} $. Furthermore, the symmetry suggests a spin representation of the form $S_{\mu} =
\mathrm{diag}\left(\sigma_{\mu},-\sigma^{*}_{\mu}  \right)$. As
$\sigma^{\dagger}_y \sigma_{\mu} \sigma_{y} = -\sigma^*_{\mu}$, we may define a
unitary operator $U = \mathbf{I} \oplus \sigma_{y}$ to rotate $S_{\mu}$ to
$\mathbf{I} \otimes \sigma_{\mu}$ for each $\mu$. Thus, for this class of
Hamiltonians, the spin Hilbert space is separable, and the effective spin
GS map is:
\begin{equation}
	\rho_{s} = \tilde{\text{Tr}}[\rho_{GS}] =  \text{Tr}_{\bar{S}}[U^{\dagger} \rho_{GS} U]
.\end{equation}

Here, $\bar{S} $
denotes the non-spin degrees of freedom following the basis transformation
achieved with $U$ rather than representing the actual non-spin degrees of
freedom.

\begin{figure}[ht]
    \includegraphics[width=\columnwidth,height=8cm]{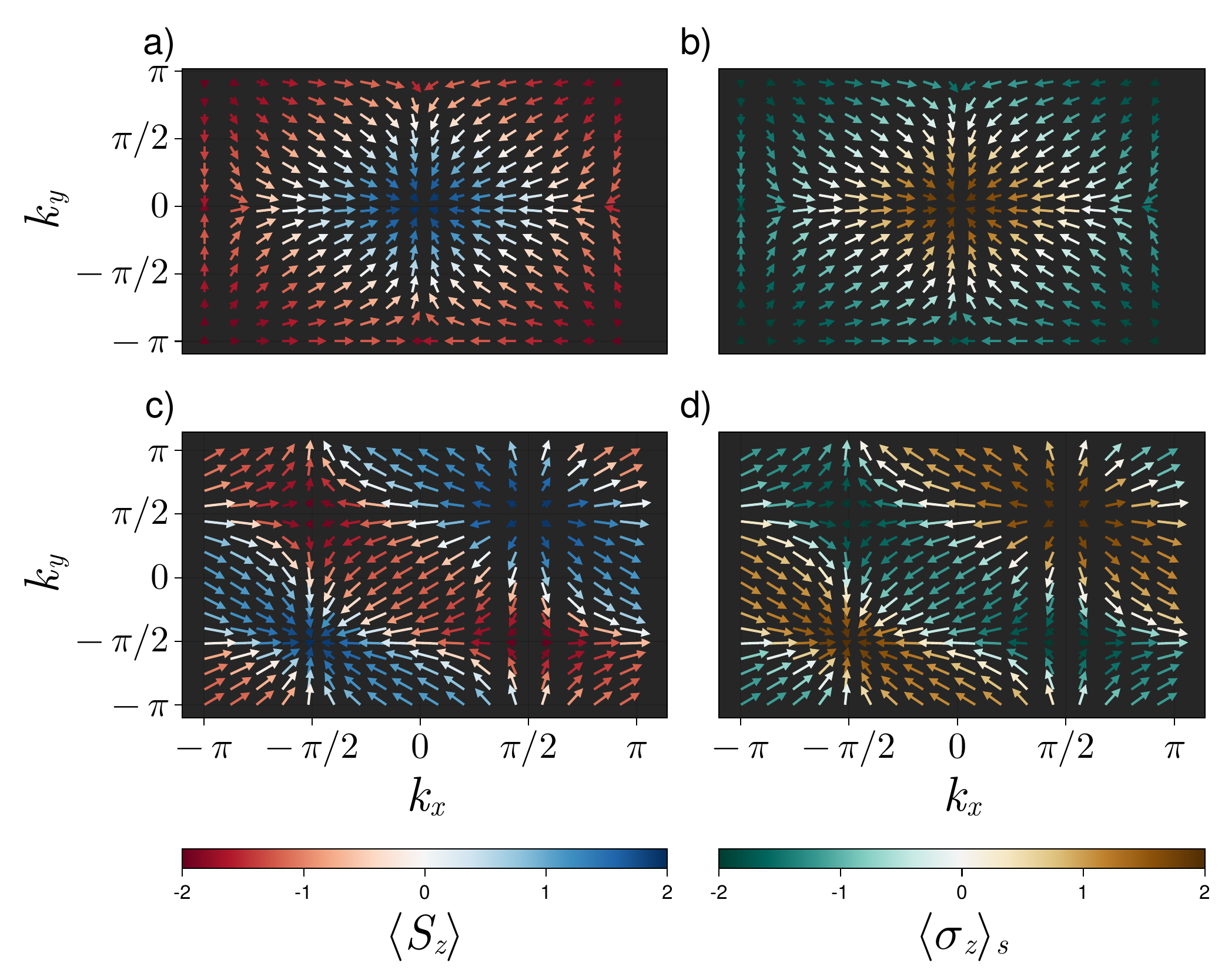}
  \vspace{-0.3cm}
  \caption{Comparison of GS spin expectation value  $\langle \boldsymbol{S} \rangle$ textures over the bulk Brillouin zone for $\boldsymbol{h} = \boldsymbol{h}_{QWZ}$ in panels a), b) and  $\boldsymbol{h} = \boldsymbol{h}_{stic}$ in c), d). Panels a),c) show $\langle \boldsymbol{S} \rangle$ computed from $\rho_{GS}$; panels b),d) the corresponding $\langle \boldsymbol{S} \rangle$ obrained from $\rho_s$. Arrow direction indicates the expectation values in the $S_x,S_y$ plane; the colour map represents the magnitude $S_z$. The arrow length is proportional to $|\langle \boldsymbol{S} \rangle|$}.
  \label{fig:1}
\end{figure}

This class Hamiltonians, with real $\boldsymbol{d}$ vector, commute with operator $\tau_{y}
\otimes \sigma_{y}$, where $\tau_{\mu}$ are the Pauli matrices in the
particle-hole Hilbert space~\cite{Varjas_2018}. They therefore block-diagonalize to the form:
\begin{equation}
	H_{BdG}(\boldsymbol{k}) = \begin{bmatrix} \left[\boldsymbol{h}(\boldsymbol{k}) + \boldsymbol{d}(\boldsymbol{k})\right] \cdot \boldsymbol{\sigma} & 0 \\ 0 & \left[\boldsymbol{h}(\boldsymbol{k}) - \boldsymbol{d}(\boldsymbol{k})\right]\cdot \boldsymbol{\sigma} \end{bmatrix}\label{eq:1}
.\end{equation}

We compute the Skyrmion number $\mc{Q}$ as in Eq.~\ref{skyrmnum} presented in the SM~\cite{sup}.

\textit{Numerical results}---To demonstrate our method through characterization
of the topological skyrmion phases, we first compare the texture over the BZ of
the spin expectation value $\langle \boldsymbol{S} (\boldsymbol{k})\rangle$ of
the GS for Eq.~\ref{BdGHam}, with the skyrmionic texture over the BZ of
$\rho_s(\boldsymbol{k})$ as shown in \cref{fig:1} for two different choices of
$\boldsymbol{h}(\boldsymbol{k})$. In \cref{fig:1} a) and \cref{fig:1} b), we
take the normal state Hamiltonian to be that of a QWZ two band Chern insulator~\cite{qi2006}, which is a two level system with $\boldsymbol{h}$-vector
$\boldsymbol{h}_{QWZ}\left(\boldsymbol{k} \right) = ( \beta \sin(k_x), \beta
\sin(k_y) ,\mu - t_q \cos(k_x) -t_q \cos(k_y))$. Here, $\mu$ defines a staggered
onsite potential; the $t_q$ are nearest-neighbour
hopping integrals; and $\beta$ a pseudo-spin orbit coupling strength. In \cref{fig:1} c)
and \cref{fig:1} d), we instead take the $\boldsymbol{h}$-vector to be that of
Sticlet~\emph{et al.}~\cite{Sticlet_2012}, or
$\boldsymbol{h}_{stic}\left(\boldsymbol{k}\right) = (\alpha \cos(k_x),\alpha
\cos(k_y), t_s \cos(k_x + k_{y})  ) $. Here, $t_s$ denotes a diagonal hopping
integral over the square lattice; and the $\alpha$ another pseudo-spin orbit
coupling.

Fig.~\ref{fig:1} confirms that the skyrmionic textures computed with the full (panels a) and c) for the two models) and with the auxiliary spin systems (corresponding panels b) and d), respectively) are identical. By construction, this equivalence is expected, but the data of fig.~\ref{fig:1} serves as the basis for our further analysis.

\textit{Observable-enriched entanglement spectrum}---Even more intriguingly, the auxiliary system produces further more intricate features, including an additional bulk-boundary correspondence arising from the spin topology. To begin the analysis, we first
perform a Fourier transform on the auxiliary spin system while also noting that
the spin operators are mutually diagonal in momenta and in position.
\begin{equation}
    \rho_s(r-r') = \sum_{r,r'} \frac{1}{2} (\mathbf{I}_{2} + \text{Tr}[\rho_{r,r'} \boldsymbol{S}] \cdot \boldsymbol{\sigma}) \ \ket{r}\bra{r'}
.\end{equation}

Here, $\rho_{r,r'}$ are the matrix elements $\braket{r|\rho|r'}$ with $r,r'$
denoting real space coordinates. Inspecting the Fourier transform, it is clear
that the procedure to reduce to the auxiliary system is to perform the standard
$N \times N \to 2 \times2$ reduction to each $r,r'$ block within the
GS projector.

\begin{figure}[ht]
    \includegraphics[width=\columnwidth,height=7cm]{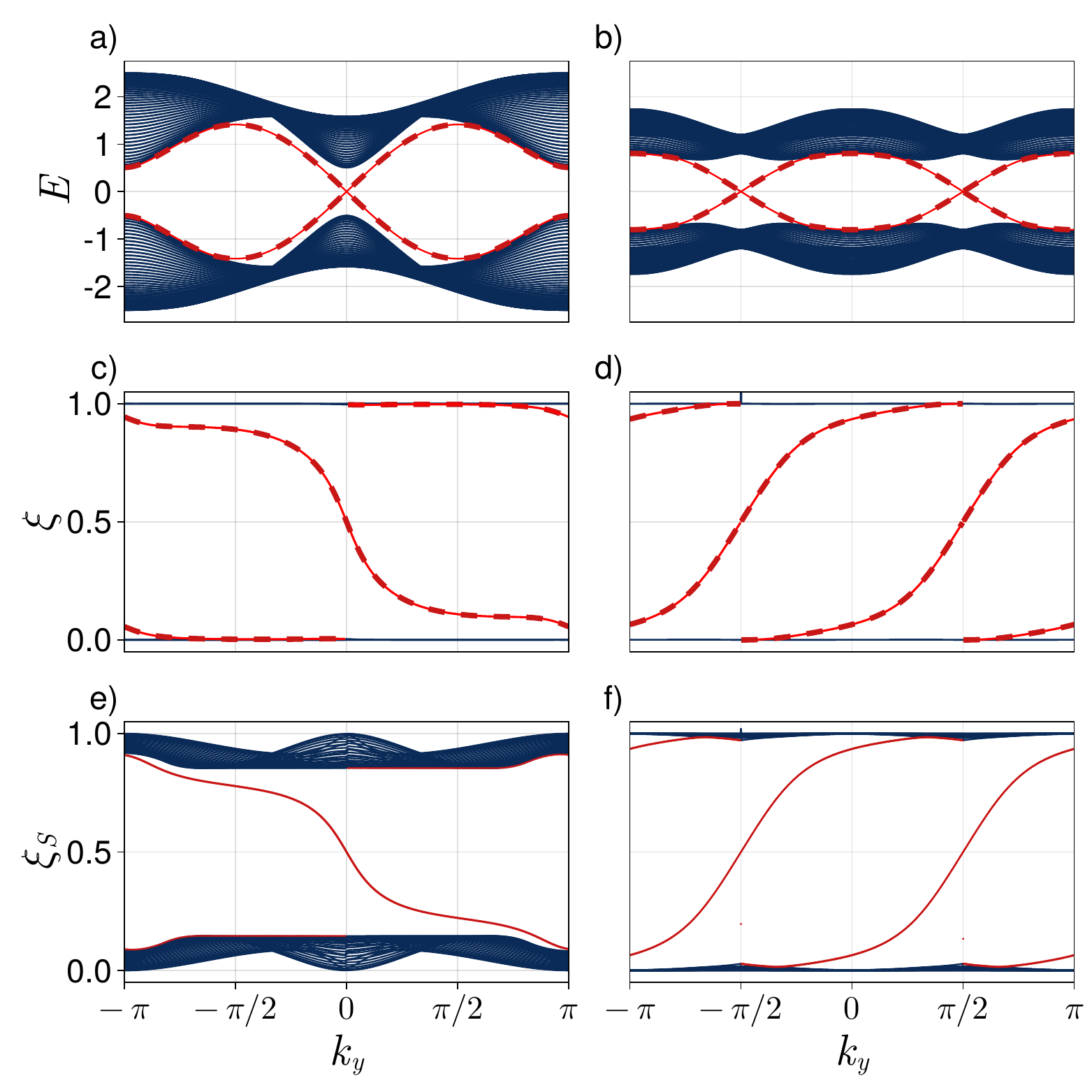}
  \vspace{-0.3cm}
  \caption{Slab energy spectra (a, b), entanglement spectra (c, d), and observable-enriched entanglement spectra (e, f) are shown for $\boldsymbol{h}_{QWZ}(\boldsymbol{k})$ (a,c,e) and $\boldsymbol{h}_{stic}(\boldsymbol{k})$ (b,d,f), respectively, for OBC in the $\hat{x}$-direction. Parameters of $\boldsymbol{h}_{QWZ}(\boldsymbol{k})$ are $\mu/t_q = 0.5,\ \Delta_0 = 1, \ \beta=1$, corresponding to $\mc{C}=2$ and $\mc{Q}=-1$. Parameters of $\boldsymbol{h}_{stic}(\boldsymbol{k})$ are $\alpha/t_S = 1\ \Delta_0 = 0.1$, corresponding to $\mc{C}=-4$ and $\mc{Q}=2$. Dashed lines in a), b), c), and d) indicate two-fold degeneracy.}
  \label{fig:2}
\end{figure}

We begin by partitioning our system in the $\hat{x}$-direction, while keeping
$k_y$ as a good quantum number, into subsystem $A$ and subsystem $B$. We then
choose the partition such that subsystem $A$ includes layers of index from $1$  to
$\frac{N_{x}}{2}$, and subsystem $B$ includes layers of index from $\frac{N_{x}}{2}+1$ up to
$N_{x}$. The partial trace over the many body groundstate corresponds to
calculating the equal time one body correlators by the method of
\cite{peschel2009}—an approach that, in the context of free fermion systems,
reduces to projecting from the GS onto the $A$ subsystem~\cite{alexandradinata2011}. We then perform an additional OEPT over
$\bar{S}$ DoFs. We present the slab energy eigenspectrum, $E$; eigenspectrum, $\xi$, of the system $A$ density matrix known as entanglement spectrum and eigenspectrum of the system $A$ density matrix after OEPT, $\xi_S$, which we denote Observable Enriched Entanglement Spectrum OEES. We compute these quantities for each of $\boldsymbol{h}_{QWZ}(\boldsymbol{k})$ and
$\boldsymbol{h}_{stic}(\boldsymbol{k})$, respectively, for representative parameter sets
in \cref{fig:2}.\par

For half-filling and $\mu/t_q = 0.5,\ \Delta_0 = 1, \ \beta=1$, the total Chern number $\mc{C}$ and Skyrmion number $\mc{Q}$
of $\boldsymbol{h}_{QWZ}(\boldsymbol{k})$ are $2$ and $-1$, respectively. We
find the slab spectrum of $\boldsymbol{h}_{QWZ}(\boldsymbol{k})$ depicted in
\cref{fig:2} a) exhibits two-fold degenerate chiral modes on each edge, as
expected from $\mc{C}$. The corresponding entanglement spectrum of the ground
state after tracing out system $B$ shown in \cref{fig:2} c) also exhibits
two-fold degenerate chiral modes as expected: chiral mode(s) in $\xi$, which connect bands at $0$ and $1$ when tuning $k_y$, are signature(s) of non-trivial topology~\cite{fidkowski2010}. However, the OEES shown in
\cref{fig:2} e) also depicts a \textit{single} chiral mode \textit{in correspondence with
$\mc{Q}$.}

For half-filling and $\alpha/t_S = 1\ \Delta_0 = 0.1$, the $\mc{C}$ and $\mc{Q}$ of
$\boldsymbol{h}_{stic}(\boldsymbol{k})$ are instead -4 and 2,
respectively. We find the slab spectrum of
$\boldsymbol{h}_{stic}(\boldsymbol{k})$ depicted in \cref{fig:2} b) exhibits a
pair of two-fold degenerate chiral modes on each edge, totaling to four chiral
modes per edge, as expected from $\mc{C}$. The corresponding entanglement
spectrum of the GS after tracing out system $B$ shown in \cref{fig:2}
d) also exhibits a pair of two-fold degenerate chiral modes as expected. The
OEES, however, shown in \cref{fig:2} f), instead exhibits
$\mc{Q}$---rather than $\mc{C}$---chiral mode(s).






\textit{Bulk-boundary correspondence for topological skyrmion phase with zero Chern number}--- The power of OEPT, however, does not lie in reproducing such results with less DoFs. We now show how it can unveil Skyrmion topology where the conventional entanglement spectrum classification suggests triviality. To this end, we break the $\tau_y \otimes \sigma_y$ symmetry by adding constant term
 $H'=\Delta_0 \tau_x \otimes \mathbf{I}_2$ to the Hamiltonian Eq.~\ref{BdGHam}, making $d_y$ complex. Such a term respects $\mathscr{C}'$ but, as shown in the SM~\cite{sup}, allows for further decoupling of $\mc{Q}$ and $\mc{C}$ yielding regions of phase space with trivial $\mc{C}$ and non-trivial $\mc{Q}$.\par


\begin{figure}[ht]
    \includegraphics[width=\columnwidth,height=7cm]{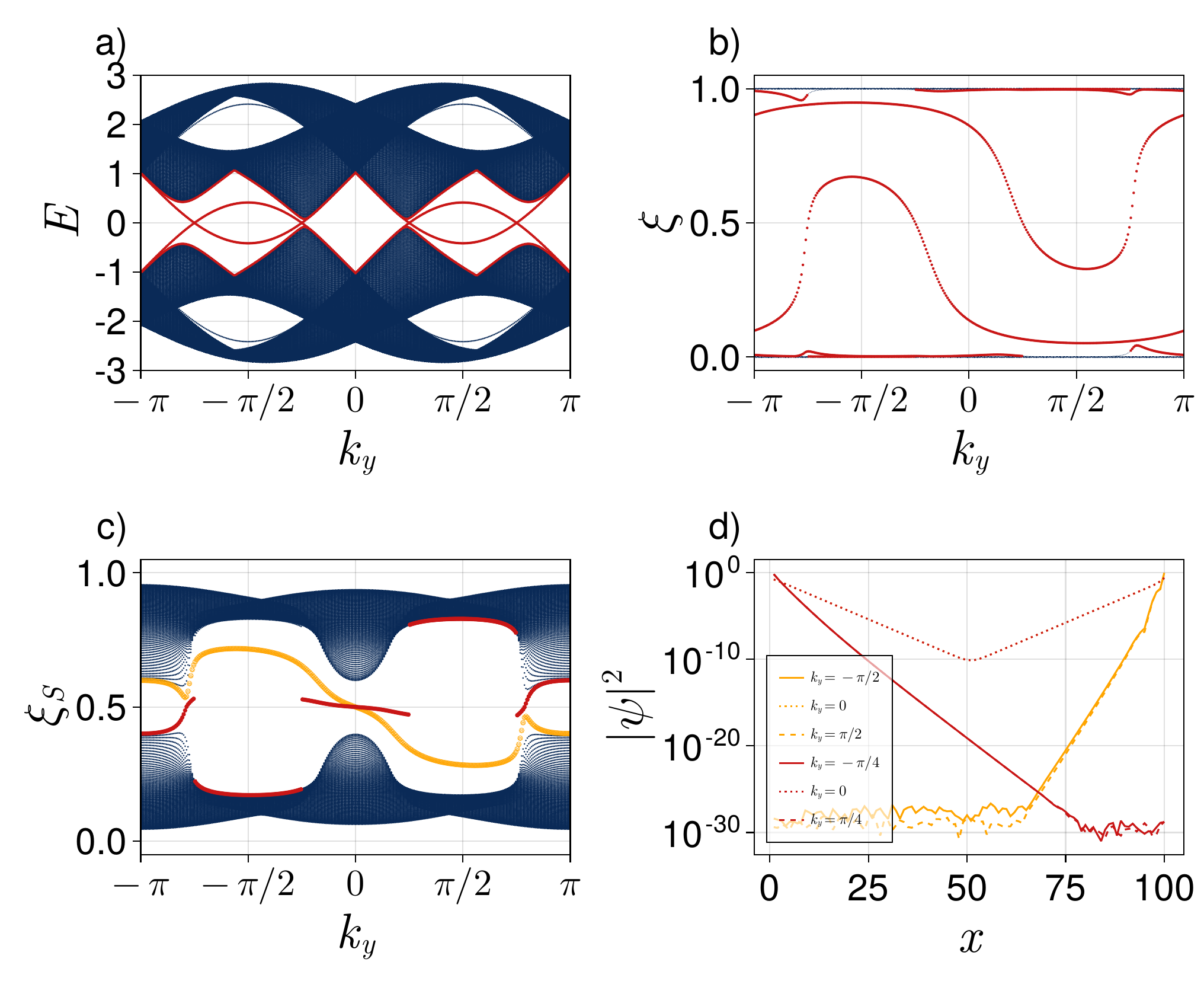}
  \vspace{-0.3cm}
  \caption{a) Slab energy spectrum for OBC in the $\hat{x}$-direction for $\mc{C}=0$, $\mc{Q}=-1$, at $m/t_q = 0.2, \Delta_0 = 1, \beta=1$, b) corresponding entanglement spectrum for cylindrical geometry, c) corresponding observable-enriched entanglement spectrum for cylindrical geometry, and d) log of probability density vs. layer index $x \in A$ for in-gap bands highlighted in red and yellow, in c). $x = 0$ corresponds to the real edge of the system; $x = 100$ corresponds to the virtual edge separating $A$ and $B$. The system is $200$ sites long.}
  \label{arcstates}
\end{figure}

The slab energy spectrum for Hamiltonian Eq.~\ref{BdGHam} including the term $H'$, with OBCs in the $\hat{x}$-direction, is shown in fig.~\ref{arcstates}~a) for a region of phase space with total Chern number $\mc{C}=0$ and Skyrmion number $\mc{Q}=-1$. Although the Chern number is zero and the system is insulating in the bulk, there are in-gap Bloch-states localized at the edge. (Localization of these edge states in the slab energy spectrum is shown in the SM~\cite{sup}.) As shown in \cref{arcstates}, these in-gap bands do not extend from the bulk valence to conduction bands or vice versa, but rather extend from the bulk valence (conduction) states into the gap, and return to the bulk valence (conduction) states. However, any value for the Fermi level that lies within the bulk gap intersects edge bands, due to the overlap of these bands in energy.

 The entanglement spectrum, $\xi$, for a virtual cut in real-space is shown in fig.~\ref{arcstates}~b). As expected for trivial total Chern number, this entanglement spectrum is trivial~\cite{fidkowski2010b}, in contrary to \cref{fig:2}, where there is spectral connectivity between $\xi=0$ and $\xi = 1$. However, the OEES, $\xi_s$, with a virtual cut in real-space, shown in fig.~\ref{arcstates}~c), exhibits a chiral mode in correspondence with the non-trivial Skyrmion number $\mc{Q}$, localized on the virtual edge as shown in fig.~\ref{arcstates}~d). Thus, the OEES correctly captures the non-trivial Skyrmion topology, where the conventional entanglement spectrum and Bloch topology fails. Furthermore, changing the sign of $\mc{Q}$ changes chirality of the edge state in the OEES~\cite{sup}. \par

  In addition, we observe a further discontinous in-gap state, red in \cref{arcstates}, localized on the real edge of the system,
 such a feature is absent in ES with OEPT in full periodic boundary
 conditions as well as with OEES with a spatial virtual cut in this geometry, which only has the expected, orange in \cref{arcstates}, chiral modes. Consequently, this implies an additional bulk-boundary correspondence, independent of the spectral contribution from Bloch states, on top of the effects seen from purely spatial cuts. This anomalous state also changes chirality with change in sign of $\mc{Q}$, further indicating that it is a consequence of the non-trivial bulk spin topology.

\begin{figure}[ht]
    \includegraphics[width=\columnwidth,height=4cm]{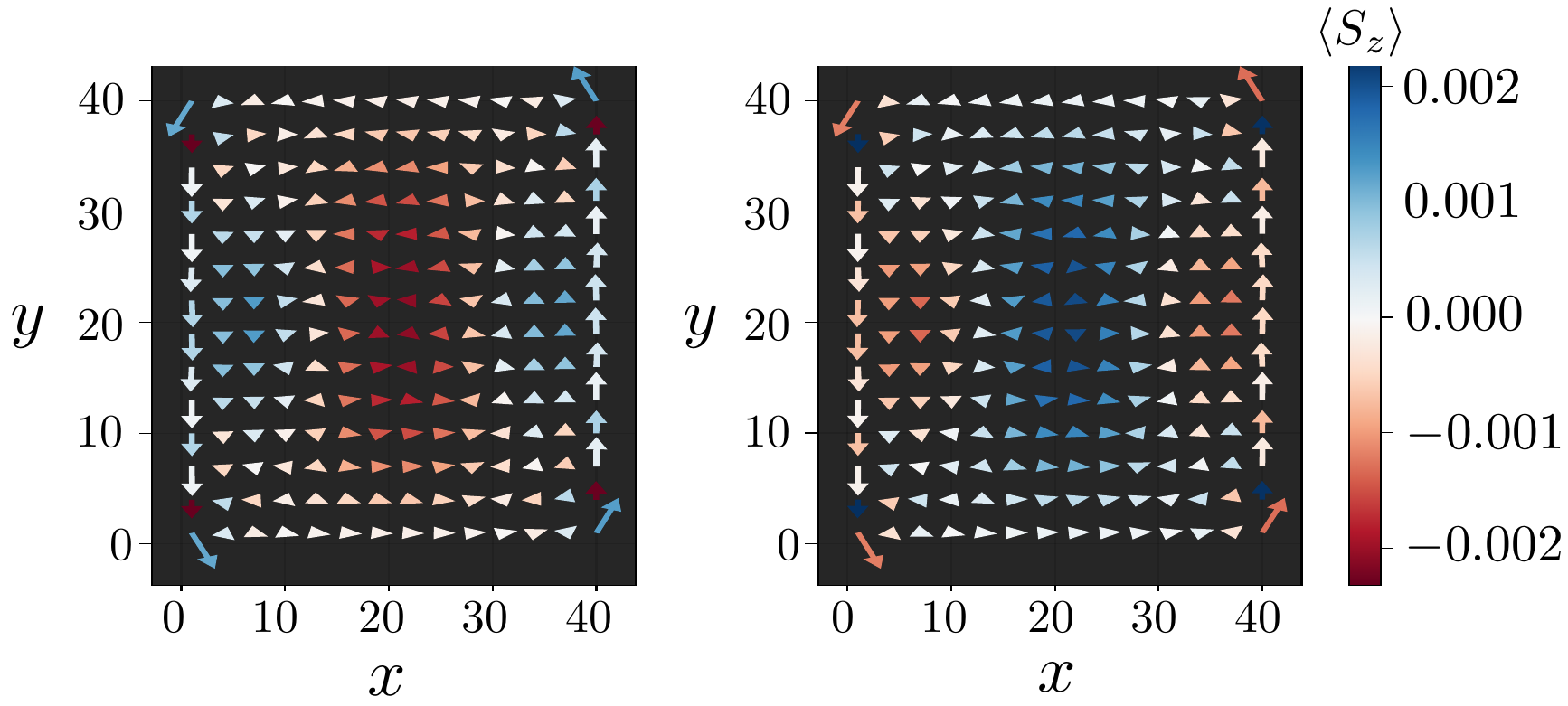}
  \vspace{-0.3cm}
  \caption{Comparison of  real-space spin expectation value $\langle \boldsymbol{S} \rangle$ texture, similar to \cref{fig:1}, for Hamiltonian Eq.~\ref{eq:1} with additional $H'$ term, for OBCs in each of the $x$- and $y$-directions. Left panel:  $\mc{Q}=1$ and $\mc{C}=0$ at  $m/t_q = -0.2, \Delta_0 = 1, \beta=1$. Right panel: $\mc{Q}=-1$ and $\mc{C}=0$ at  $m/t_q = 0.2, \Delta_0 = 1, \beta=1$}
  \label{fig4}
\end{figure}

 Supplementary to the OEES, upon opening boundary conditions we see a real space chiral spin texture on the boundary of the system as shown in \cref{fig4}. Change in sign of $\mc{Q}$ also corresponds to change in handedness of $\langle S_z \rangle$ of the spin texture in the full model, as shown in fig.~\ref{fig4}. Such effects show how the non trivial OEES manifests itself on the real boundary of the system. \par


\textit{Discussion and conclusion}---We have introduced the concept of observable enriched entanglement. This is based on a generalisation of the partial trace operation to obtain an  entanglement spectrum. The result is an enriched partial trace which preserves the values of the selected observables. Applied to the ground state (GS) this produces an auxiliary density matrix, $\rho_s$, with DoF only associated with the observable. Remarkably, we find that the $\rho_s$ reproduces the entanglement spectra corresponding to non-trivial spin topology of the full GS with non-zero Chern number. In addition, however, we could reveal that $\rho_s$ contains further information on bulk-boundary correspondence of the observable even for zero Chern number, where the standard entanglement spectrum is inconclusive. We demonstrated these features on different models with Chern and Skyrmion numbers in the context of revealing edge polarisation textures bound to the Skyrmion number. Future work will more broadly characterise other entanglement measures with established expressions in terms of reduced density matrices (e.g., von Neumann entropy~\cite{eisert2010}), but computed instead from OE reduced density matrices.

\begin{acknowledgments}

\textit{Acknowledgements}---We thank  R. F. Calderon, G. F. Lange, S. Liu, M. Pacholski, A. Pal, and D. Varjas for fruitful discussions. This research was supported in part by the National
Science Foundation under Grants No.NSF PHY-1748958 and PHY-2309135, and
undertaken in part at Aspen Center for Physics, which is supported by National
Science Foundation grant PHY-2210452. The work presented in this paper is theoretical. No data were produced, and supporting research data are not required.

\end{acknowledgments}

\bibliography{p1bib.bib}

\clearpage

\title{Supplemental material for ``Observable Enriched Entanglement''}
\makeatletter
\renewcommand{\theequation}{S\arabic{equation}}
\renewcommand{\thefigure}{S\arabic{figure}}
\renewcommand{\thesection}{S\arabic{section}}
\setcounter{equation}{0}
\setcounter{section}{0}
\onecolumngrid
\begin{center}
  \textbf{\large Supplemental material for ``Observable-enriched entanglement'}\\[.2cm]
  Joe H. Winter$^{1,2,3}$, Reyhan Ay$^{1,2,4}$, Bernd Braunecker$^{3}$ and Ashley M. Cook$^{1,2,*}$\\[.1cm]
  {\itshape ${}^1$Max Planck Institute for Chemical Physics of Solids, Nöthnitzer Strasse 40, 01187 Dresden, Germany\\
  ${}^2$Max Planck Institute for the Physics of Complex Systems, Nöthnitzer Strasse 38, 01187 Dresden, Germany\\
  ${}^3$SUPA, School of Physics and Astronomy, University of St.\ Andrews, North Haugh, St.\ Andrews KY16 9SS, UK\\
  ${}^4$Izmir Institute of Technology, Gülbahçe Kampüsü, 35430 Urla Izmir, Türkiye\\}
  ${}^*$Electronic address: cooka@pks.mpg.de\\
(Dated: \today)\\[1cm]
\end{center}

\section{Analytic Calculation of Topological Invariants}

 In this section, we extend the discussion of the bulk topological invariants for the Hamiltonians presented in the main text. We begin by utilising the $\tau_{y} \otimes \sigma_{y}$ symmetry, where $\tau_{\mu}$ are the Pauli matrices in the particle-hole Hilbert space. This allows reduction of the Hamiltonian to a simple block diagonalization:

\begin{equation}
	H_{BdG}(\boldsymbol{k}) = \begin{bmatrix} (\boldsymbol{h} + \boldsymbol{d})(\boldsymbol{k}) \cdot \boldsymbol{\sigma} & 0 \\ 0 & (\boldsymbol{h} - \boldsymbol{d})(\boldsymbol{k}) \cdot \boldsymbol{\sigma} \end{bmatrix}\label{eq:1}
.\end{equation}

Furthermore, and crucially, the spins still remain separable with some directions modified:

\begin{equation}
    S_x \to \boldsymbol{I} \otimes -\sigma_{x},\ S_y \to \boldsymbol{I} \otimes -\sigma_{y},\ S_z \to \boldsymbol{I} \otimes \sigma_{z}
.\end{equation}

The modification is effectively a change in handedness of the Bloch sphere, so any winding calculated with the tuple $(-\sigma_{x},-\sigma_{y},\sigma_{z})$ has opposite sign to our original set of Pauli matrices.\par

With these lemmas, we can now perform a topological band theoretical analysis of these models. Firstly, we utilise the block diagonal form in \cref{eq:1} to calculate the ground-state topology of this Hamiltonian. Given each block possesses two bands and is, individually, $\mathscr{C}'$ symmetric, the total Chern number of the four-band system is the sum of the Chern numbers of the two-band blocks. This therefore reduces to the sum of Skyrmion numbers of the vectors $\boldsymbol{h}+\boldsymbol{d}, \boldsymbol{h}-\boldsymbol{d}$.

\begin{equation}
    \mathcal{C} = \mathcal{Q}[\boldsymbol{h} + \boldsymbol{d}] + \mathcal{Q}[\boldsymbol{h} - \boldsymbol{d}]
,\end{equation}

where $\mathcal{Q}[\boldsymbol{v(k)}]$ denotes the skyrmion number computed as the winding of vector $ \boldsymbol{v}$ in the Brillouin zone. We take the partial trace in this block-diagonal basis, noting we have a different handed winding, which gives:

\begin{equation}
    \rho_{S} = \rho^{+}_{GS} + \rho^{-}_{GS},
    \end{equation}

where $\rho^{\pm}_{GS}$ denotes the rank $1$ ground-state projector of $(\boldsymbol{h} \pm \boldsymbol{d}) \cdot \sigma$, respectively. This now allows us to very simply calculate the Skyrmion number for our four-band system, which is the Chern number/Skyrmion number of this effective two-band system. As such, taking into account our rotation of spin operators where $ \braket{\sigma_{x,y}}_S = -\braket{S_{x,y}} $, we find that the four-band Skyrmion number is equivelent to
 :

\begin{equation}
	\mathcal{Q}_{tot} = - \mathcal{Q}\left[ \frac{(\boldsymbol{h} + \boldsymbol{d})}{|\boldsymbol{h}+\boldsymbol{d}|} + \frac{(\boldsymbol{h}-\boldsymbol{d})}{|\boldsymbol{h}-\boldsymbol{d}|} \right]
.\end{equation}


Now, the integrand for the Skyrmion number is non-linear and, typically, we cannot write a simple formula in terms of a linear combination of winding numbers of the individual $\boldsymbol{h} \pm \boldsymbol{d}$ vectors. However, we are saved by the fact that we are looking a topological quantity. First, assume we compute the winding number of:

\begin{equation}
	\mathcal{Q}_{tot}(\alpha) = - \mathcal{Q}\left[  \frac{(\boldsymbol{h} + \boldsymbol{d})}{|\boldsymbol{h}+\boldsymbol{d}|} + \alpha \frac{(\boldsymbol{h}-\boldsymbol{d})}{|\boldsymbol{h}-\boldsymbol{d}|}\right], \ \ \alpha \in [0,1].
 \end{equation}

If we first assume $\alpha = 0$, this is simply the winding of the normalised $\boldsymbol{h}+\boldsymbol{d}$ vector. Now, provided $\alpha \in [0,1)$, the vector $\boldsymbol{h}-\boldsymbol{d}$ cannot cancel the vector $\boldsymbol{h} + \boldsymbol{d}$, as its magnitude is always less than one. Consequently, we remain in the same topological class. Therefore, when $\alpha=1$, we either remain in the same topological class or the total vector passes through zero in magnitude and the integral is ill-defined. Therefore, employing this effective spin ground state, we have proved the statement previously verified only numerically:

\begin{equation}
    \mathcal{C} = -2 \mathcal{Q}
.\end{equation}

$\mathcal{Q}[\boldsymbol{h}+\boldsymbol{d}] = \mathcal{Q}[\boldsymbol{h}-\boldsymbol{d}]$ for the case where we have $\tau_y \otimes \sigma_y$ symmetry.

\section{Phase Diagram for Complex $\boldsymbol{d}$-vector}

\begin{figure}[!htb]
    \includegraphics[width=8cm,height=6cm]{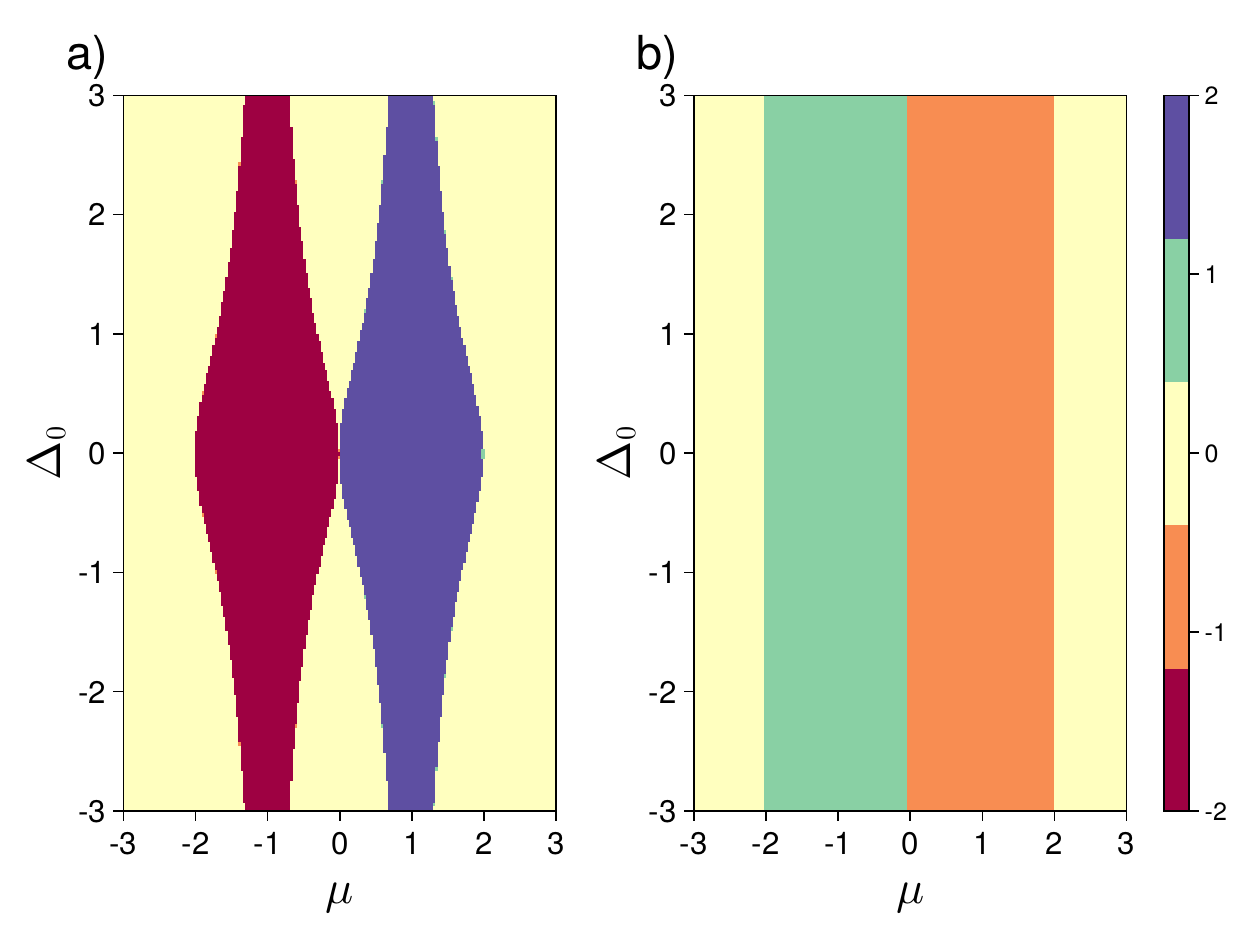}
  \vspace{-0.3cm}
  \caption{Topological phase diagrams for Hamiltonian Eq.~3 with additional $H'$ term corresponding to results for Figs.~3 and 4 in the main text: a) total Chern number $\mc{C}$ and b) skyrmion number $\mc{Q}$, for half-filling, vs. normal state mass parameter $\mu/t_q$ and pairing strength $\Delta_0$ with pseudo-spin orbit coupling $\beta = 1$.}
  \label{S8}
\end{figure}

Here, we present phase diagrams for the Hamiltonian Eq.~\ref{BdGHam} with additional $H'$ term in fig.~\ref{S8}. As shown in fig.~\ref{S8} a), regions with non-trivial total Chern number for half-filling narrow with increasing magnitude of pairing strength $\Delta_0$, while regions with non-trivial skyrmion number are independent of $\Delta_0$ as shown in fig.~\ref{S8} b). As result, finite $\Delta_0$ yields regions of phase space with trivial total Chern number and non-trivial skyrmion number. In addition, type-II topological phase transitions are realized for $\mu=0$ and $\pm 2$, across which the skyrmion number changes from one integer value to another without the closing of the minimum direct bulk energy gap. Instead, the skyrmion number changes as the minimum magnitude of the ground state spin expectation value over the Brillouin zone becomes zero.

\pagebreak

\section{OBC Characterisation for Complex $\boldsymbol{d}$-vector with $\mathcal{Q}=+1$}

\begin{figure}[!htb]
    \includegraphics[width=8cm,height=6cm]{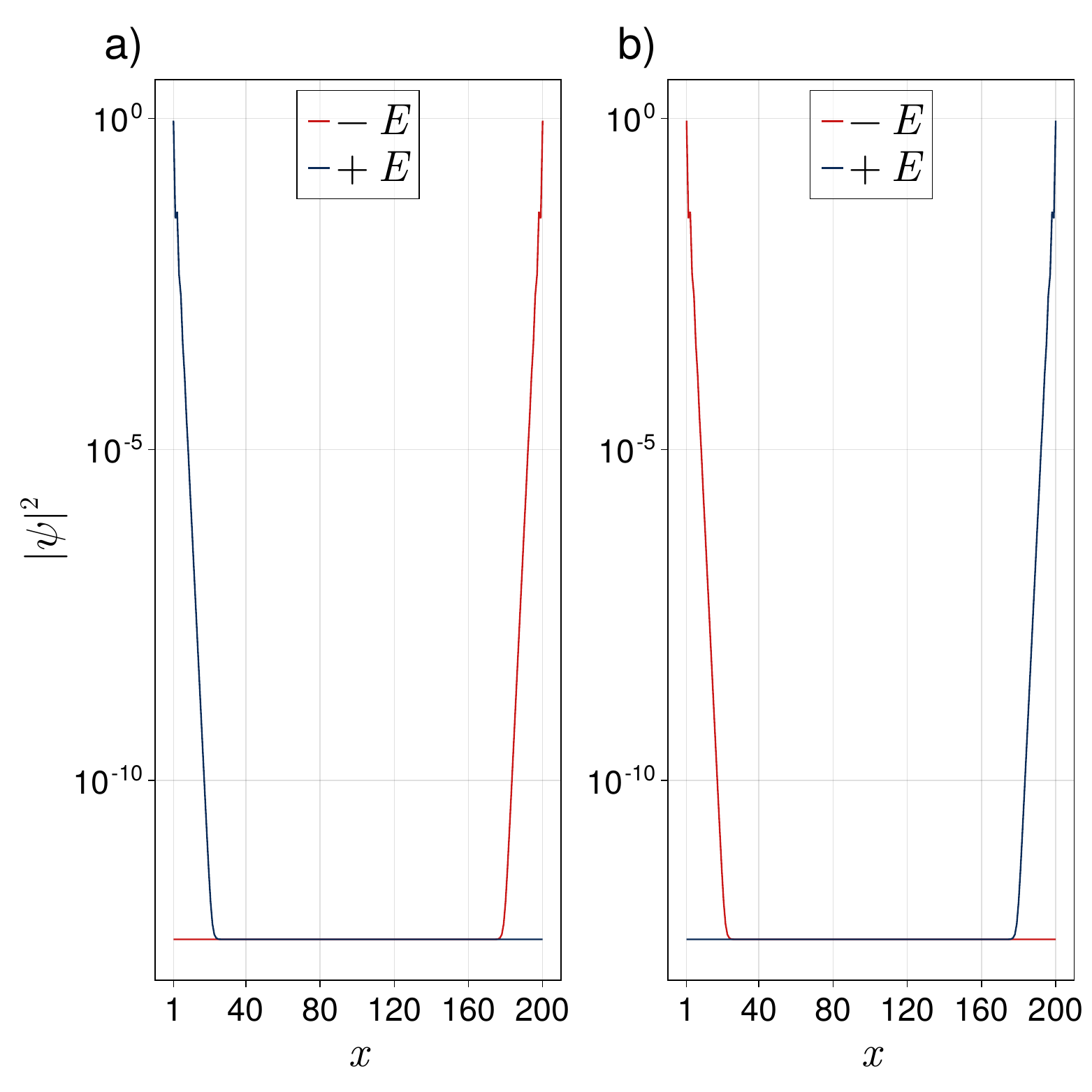}
  \vspace{-0.3cm}
  \caption{Probability density vs. layer index for the two in-gap edge states shown in fig.~3 a) (highlighted in red) at a) $k_y=-\pi/2$ and b) $k_y = \pi/2$, respectively.}
\end{figure}

\begin{figure}[!htb]
    \includegraphics[width=8cm,height=8cm]{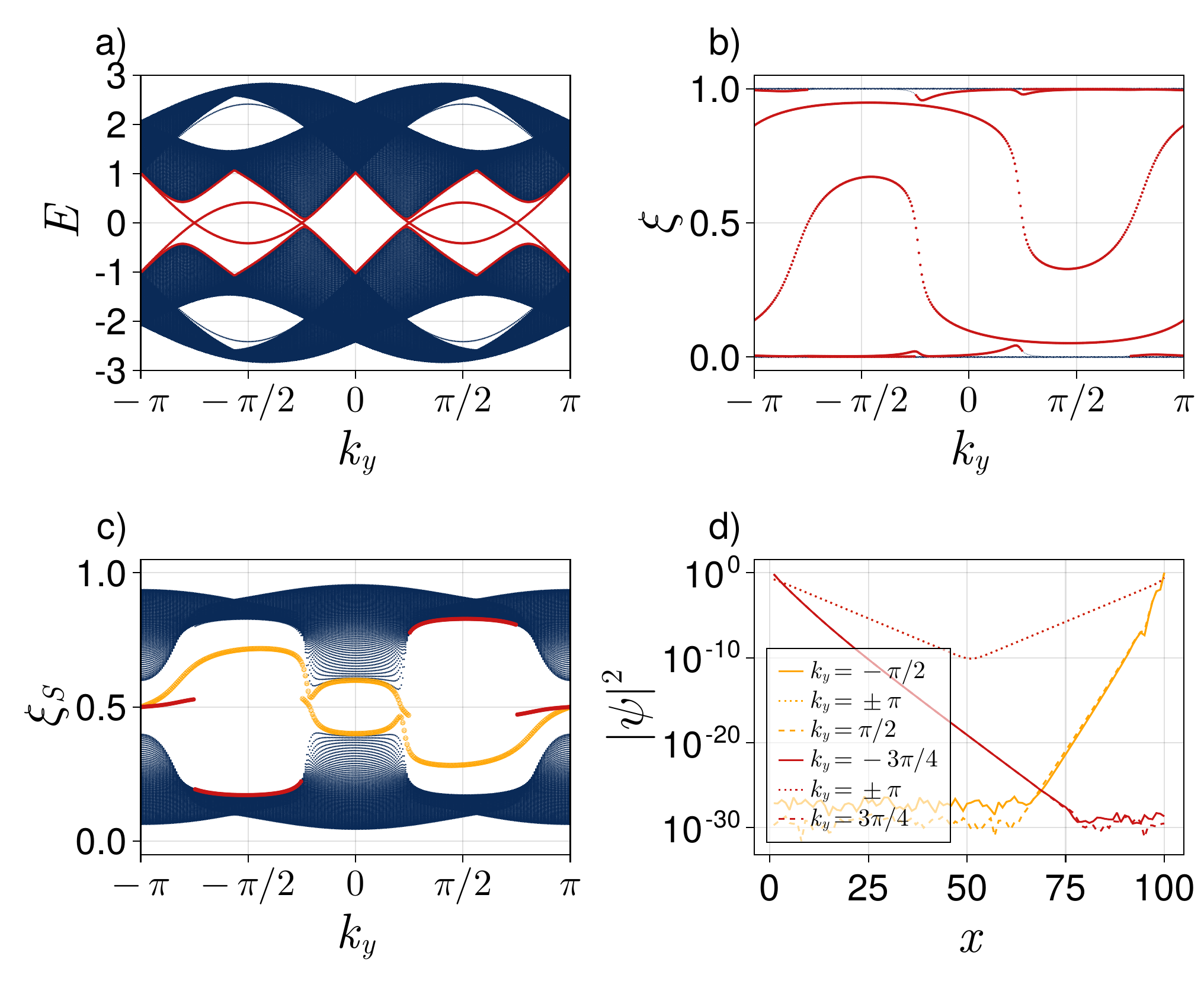}
  \vspace{-0.3cm}
  \caption{a) Slab energy spectrum for OBC in the $\hat{x}$-direction for $\mc{Q}=1$, $\mc{C}=0$ and parameter set $\mu / t_q = -0.2, \Delta_0 = 1, \beta = 1$, b) corresponding entanglement spectrum for cylindrical geometry and c) observable-enriched entanglement spectrum for cylindrical geometry, and d) log of probability density vs. layer index $x$ for in-gap bands highlighted in red and yellow, in c).}
  \label{S7}
\end{figure}

In fig.~\ref{S7}, we present the complement to fig.~3 but for $\mc{Q}=1$ and $\mc{C}=0$,  and parameter set $\mu / t_q = -0.2, \Delta_0 = 1, \beta = 1$. We see that there is a difference in chirality in panel c) of both the expected chiral state, in orange, and the additional state living on the real edge, red, relative to the corresponding in-gap states in fig.~3 c) in the main text. This is further evidence that these features are in correspondence with the Skyrmion number of the bulk.

\section{Periodic Boundary Conditions OEES for Complex $\boldsymbol{d}$-vector with $\mathcal{Q}=-1$}

\begin{figure}[!htb]
    \includegraphics[width=10cm,height=10cm]{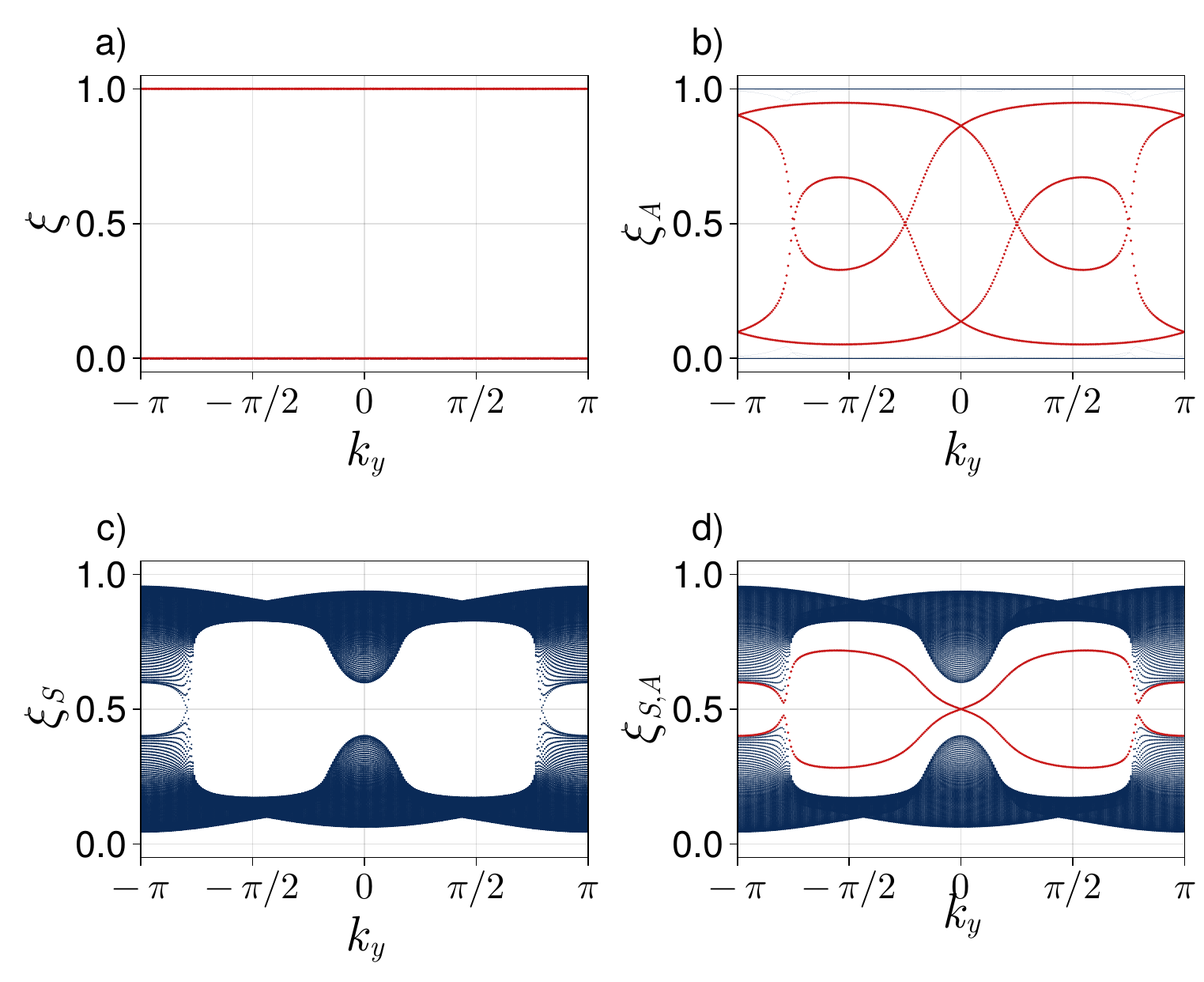}
  \vspace{-0.3cm}
  \caption{Various spectra calculated for the model with Complex $\boldsymbol{d}$ vector at parameters  $\mu / t_q = -0.2, \Delta_0 = 1, \beta = 1$, $\mc{Q}=-1$, $\mc{C}=0$ in full periodic boundary conditions. Here panel a) is the ground state entanglement spectrum $\xi$; b) the ground state entanglement spectrum with virtual cut on the torus of sites $x=101$ to $x=200$ inclusive, $\xi_A$; c) ground state entanglment spectrum after OEPT, $\xi_S$; d) ground state density matrix with both OEPT and virtual cut on torus, $\xi_{S,A}$.}
  \label{S9}
\end{figure}

Fig.~\ref{S9} displays results on various entanglement spectra in a cylindrical geometry -- with open boundary conditions in one direction and periodic boundary conditions in the other -- to a toroidal geometry -- with full periodic boundary conditions. We see in panel a) the eigenspectrum, $\xi$ of the groundstate density matrix, which is composed of pure states only. Panel b) shows the spectrum $\xi_A$, which is the eigenspectrum after a virtual cut on the last $100$ sites of the torus, forming two entanglement edges. We see the states are trivial as expected, as they do not connect the spectrum at $\xi_A = 0$ and $\xi_A = 1$. Panel c) shows the ground state with only the OEPT applied to it, $\xi_S$: edge states are absent, in contrast to the entanglement spectrum presented in fig.~3 c). Finally, d) shows the $\xi_S$ spectrum after a real space virtual cut of the type in b), $\xi_{S,A}$. We see two crossing chiral states of the type presented in orange in fig.~3 d).

\end{document}